\documentclass[12pt]{article}
\usepackage{epsfig}

\begin{document}
\title{Cosmic String Loops Collapsing to Black Holes}
\author{R.N. Hansen\thanks{Electronic address: rnh@fysik.ou.dk}, \,
M. Christensen\thanks{Electronic address: inane@fysik.sdu.dk} \, and
A.L. Larsen\thanks{Electronic address:  all@fysik.ou.dk}}
\date{\today}
\maketitle
\noindent
\centerline{\em Physics Department, University of Odense, }\\
\centerline{\em Campusvej 55, 5230 Odense M, Denmark}
\vskip 2cm

\begin{abstract}
\baselineskip=1.5em
We examine the question of collapse of Turok's
two-parameter family of cosmic
strings. We
first perform a classification of the strings according to the
specific time(s) at which the
minimal string size is reached during one period. We then obtain an exact
analytical expression
for the probability of collapse to black holes for the Turok strings. Our
result has the same
general behavior as previously obtained in the literature but we find, in
addition, a
numerical prefactor that changes the result by three orders of magnitude.
Finally we show that our
careful computation of the prefactor  helps to
understand the discrepancy between previously obtained results and, in
particular,
that for ``large" values of $G\mu$, there may not even be a discrepancy. We
also give a
simple physical argument that can immediately rule out some of the previously
obtained results.
\end{abstract}

\newpage
\section{Introduction}
\label{sec1}
\setcounter{equation}{0}

According to the hoop-conjecture \cite{hoop}, a cosmic string that
contracts to a size smaller
than its Schwarzschild radius will collapse and form a black hole. This
process is of particular
interest and importance in connection with primordial black holes, Hawking
radiation, high energy
cosmic gamma bursts etc.

In a network of cosmic strings, only a very small fraction, $f$, of strings
are expected to
collapse to black holes. Many attempts have been made to obtain a value
for $f$ (see for instance
\cite{pol}-\cite{fort}),
but the results deviate wildly.

Interestingly enough, only in one of the pioneering papers on the subject,
the one by Polnarev and
Zembowicz \cite{pol}, is the derivation of
$f$ based on exact analytical expressions for the cosmic strings involved
in the process. In all
other discussions the derivation of $f$ is based on linearized expressions
for the string
configurations (for
instance \cite{garriga,maria}), rather general arguments and estimates (for
instance
\cite{hawking,honma}), observational data concerning high energy cosmic
rays (for instance
\cite{gates}-\cite{ubi}) or the derivation is purely numerical (for
instance \cite{xin,casper}). And even in Ref. \cite{pol}, the final
computation of $f$ is actually
numerical, although it could in fact have been performed analytically.

In the present paper we consider the question of collapse for the
analytical two-parameter family
of strings introduced by Turok \cite{turok}. This is the same family of
strings that was
considered in \cite{pol}. However, simple explicit examples
show that the results obtained in \cite{pol} are correct only in part of
the parameter-space. We will show
that some essential points were missed in Ref. \cite{pol}, and that the
results obtained there are
not completely correct.

First of all, we make a classification of the string configurations
according to their general behavior during one period of oscillation.
Together with some simple
explicit examples, this analysis reveals that the corresponding result
obtained in \cite{pol} is in
fact only correct in approximately half of the two-dimensional
parameter-space. Secondly, we then
derive the exact analytical expression for the probability  $f$ of string
collapse to black
holes. Our result for $f$ agrees partly with that of Ref. \cite{pol} in the
sense that
$f\propto(G\mu)^{5/2}$, where $\mu$ is the string tension and $G$ is
Newtons constant. However, we
find a numerical prefactor in the relation, expressed in terms of Euler's
gamma-function, which is
of the order $2000$. For comparison, in Ref. \cite{pol} the prefactor was
found (numerically) to be of the order $1$. Finally we show that our
careful computation of the prefactor  helps to
understand the discrepancy between previously obtained results and, in
particular,
that for ``large" values of $G\mu$, there may not even be a discrepancy. We
also give a
simple physical argument that can immediately rule out some of the previously
obtained results.

The paper is organized as follows. In Section \ref{sec2}, we
introduce the string
configurations of Turok
\cite{turok} and discuss the results obtained by Polnarev and Zembowicz
\cite{pol}. In Section \ref{sec3},
we give the precise
definition of the minimal string size $R$, and we make a complete
classification of the
two-dimensional parameter-space. The classification is in terms
of the specific time(s) at which
the value $R$ comes out for a particular string configuration during one
period of ``oscillation".
In Section \ref{sec4}, we derive an exact analytical
expression for the probability
of string collapse to
black holes and we give the approximate result in the physically realistic
limit $G\mu<<1$.
Finally in Section \ref{sec5}, we argue that our
careful computation of the prefactor  helps to
understand the discrepancy between previously obtained results, and we give our
concluding remarks.  Some of the details
of the computations used
in sections \ref{sec3} and \ref{sec4} are presented in the Appendix. \\

\section{Two-Parameter Turok-Strings}
\label{sec2}
\setcounter{equation}{0}

The string equations of motion in flat Minkowski space take the form:
\begin{equation}
\ddot X^{\mu} - X^{''\mu} = 0
\label{waveeq}
\end{equation}
supplemented by the constraints:
\begin{equation}
\eta_{\mu\nu} \dot X^{\mu}X^{'\nu} = \eta_{\mu\nu}\left(\dot X^{\mu} \dot
X^{\nu} + X^{'\mu}  X^{'\nu}\right) = 0
\label{gaugefix}
\end{equation}
It is convenient to take $X^{0}= A\tau$, where $A$ is a constant with
dimension of length.
Then Eqs. (\ref{waveeq})-(\ref{gaugefix}) become:
\begin{eqnarray}
\ddot{\!\! \vec{X}} & = & \vec{X}^{''} \nonumber\\
\dot{\! \vec{X}}\cdot\vec{X}^{'} & = & 0 \label{neweom} \\
\dot{\vec{X}^2} + \vec{X}^{'2} & = & A^2 \nonumber
\end{eqnarray}
A two-parameter solution to Eqs. (\ref{neweom}) was first introduced by Turok
\cite{turok}:
\begin{eqnarray}
X(\tau,\sigma) & = & \frac{A}{2}~\left[{(1-\alpha)\sin (\sigma -\tau) +
\frac{\alpha}{3}\sin
3 (\sigma - \tau)+ \sin (\sigma +\tau) }\right] \nonumber\\
Y(\tau,\sigma) & = & \frac{A}{2}~\left[{(1-\alpha)\cos (\sigma -\tau) +
\frac{\alpha}{3}\cos
3 (\sigma - \tau)+ (1-2 \beta )
\cos (\sigma +\tau) }\right] \nonumber \\
Z(\tau,\sigma) & = & \frac{A}{2}~\left[{2\sqrt{\alpha(1-\alpha)}\cos (\sigma -
\tau) +
2\sqrt{\beta(1-\beta)}\cos (\sigma + \tau)  }\right] \label{turokstring}
\end{eqnarray}
where $\alpha\in{[0;1]}$, $\;\beta\in{[0;1]}$. This family of solutions
generalizes the solutions
considered in Ref. \cite{kibble}.

The total mass-energy of the
Turok-string is:\
\begin{equation}
\mbox{Energy} = -\int{ d^{3}\vec{X}~T^0\:_{0}  }
\end{equation}
where:\
\begin{equation}
\mbox{T}_{\mu\nu} \equiv \frac{2}{\sqrt{-g}} \frac{\delta S}{\delta
g^{\mu\nu}}=
\mu \int d\tau d\sigma\; \eta_{\mu\lambda} \eta_{\nu\delta}
\left(\dot X^{\lambda} \dot X^{\delta} - X^{'\lambda}  X^{'\delta}\right)
\delta^{4}\left(X - X(\tau, \sigma)\right)
\end{equation}
It follows that:
\begin{equation}
\mbox{Energy} = 2\pi A\mu
\label{stringenergy}
\end{equation}
which (by construction) is independent of the parameters ($\alpha, \beta$).
Similarly, one finds for the momenta:
\begin{equation}
\mbox{P}^i = -\int{d^{3} \vec{X}~ T^{i}\;_{0}} = 0
\end{equation}
In fact, the string center of mass is located at $\vec{X}_{cm} = (0,0,0) $ at
all times,
and the strings are symmetric under reflection in origo.\
The Schwarzschild radius corresponding to the energy (\ref{stringenergy}) is
\begin{equation}
R_S = 4\pi AG\mu
\end{equation}
For realistic cosmic strings, the dimensionless parameter $G\mu$ is quite
small \cite{vilenkin,shellard}, say $G\mu \sim 10^{-6}$. It is then clear
that a
Turok-string (\ref{turokstring})
will typically be far outside its Schwarzschild radius (as follows since
typically $X \sim A,
\;Y \sim A,\; Z\sim A$). However, for certain particular values $(\alpha,
\beta)$,
a string might at some instant during its evolution be completely inside
its Schwarzschild
radius. Such a string will collapse and eventually form a black hole,
according to the socalled hoop-conjecture \cite{hoop}. To determine whether
this happens
or not, we must first find the minimal 3-sphere, that completely encloses the
string, as
a function of time. After minimization over time, we then get the minimal
3-sphere that can ever
enclose the whole string.  Let the radius of this sphere be
$R$ (it will be defined more stringently in the next
section). Then the condition for collapse is:
\begin{equation}
R \leq R_S
\end{equation}
That is to say,
\begin{equation}
\frac{R}{A} \leq 4\pi G\mu
\label{condition}
\end{equation}

In a pioneering paper, Polnarev and Zembowicz \cite{pol} considered,
among other things, the question of collapse of the Turok-strings
(\ref{turokstring}).
In connection with the minimal string size $R$ they found:
\begin{itemize}
\item The strings have their minimal size $R$ at
\begin{equation}
\tau = \frac{\pi}{2}
\label{polz1}
\end{equation}
\item For generic parameters $(\alpha, \beta)$:
\begin{equation}
\frac {R^2}{A^2} = \left(\sqrt{\alpha(1-\alpha)} -
\sqrt{\beta(1-\beta)}\;\right)^2 +
\left(\frac{\alpha}{3}-\beta\right)^2
\label{polz2}
\end{equation}
\end{itemize}
The result (\ref{polz1}) would be expected
for a string experiencing a monopole-like
oscillation, i.e. starting from maximal size at $\tau = 0$, then
contracting isotropically to minimal
size at $\tau = \pi/2$, and then re-expanding to its original maximal size at
$\tau = \pi $. As for the result (\ref{polz2}) it was simply stated \cite{pol}
without proof or derivation.

In Sections 3,4 we shall show that
(\ref{polz1}) and (\ref{polz2}) are not completely correct. In fact, they
are correct
in  part of the two-dimensional parameter-space but incorrect in other
parts. A simple
example
showing that  (\ref{polz2}) cannot be correct, is provided by
the case $(\alpha , \beta) = (3/4,1/4)$. In that case
the result (\ref{polz2}) would actually give $R = 0$,
which would mean that the whole
string had collapsed to a point at some instant. This would imply that at
some instant
$X = Y = Z = 0$, but that is certainly impossible for $(\alpha , \beta) =
(3/4,1/4)$. The problem is
that (\ref{polz2}) gives the minimal distance
from origo to the
string, but it is the maximal distance which is relevant for the minimal
3-sphere.\\

\section{Minimal String Size and Classification}
\label{sec3}
\setcounter{equation}{0}

For a given pair of parameters ($\alpha , \beta$), we define the minimal
string size $R^2$ as the (square of the) radius of the minimal 3-sphere
that can ever  enclose the string completely. More precisely:
\begin{equation}
{R}^2 \equiv
\begin{array}{c}
\mbox{Minimum}\\
\tau\in\left[0; \pi\right]
\end{array}
\left[
\begin{array}{cc}
\mbox{Maximum}\\
\sigma\in\left[0; \pi\right]
\end{array}
\left(R^2\left(\tau, \sigma\right)\right)\right]
\label{defr}
\end{equation}
where:
\begin{equation}
{R}^{2}\left(\tau,\,\sigma\right) = X^2(\tau, \sigma) + Y^2(\tau, \sigma) +
Z^2(\tau, \sigma)
\label{defdist}
\end{equation}
as obtained using Eq. (\ref{turokstring}).
Thus, for a fixed time $\tau$, we first compute the maximal distance from
origo (i.e. the string center of mass) to the string. This gives the minimal
3-sphere that encloses the string at that particular instant. We then minimize
this maximal distance over all times. This gives altogether the minimal
string size
$R^2$. And this is obviously the quantity that must be compared with the
(square
of the) Schwarzschild radius $R^2_S$. Notice that we need only maximize over
$\sigma\in\left[0; \pi\right]$ and minimize over $\tau\in\left[0; \pi\right]$
in Eq. (\ref{defr}). This is due to the reflection symmetry and time
periodicity of
the Turok-strings (\ref{turokstring}).\\

We now outline the computation of $R^2$. The details can be found in the
Appendix,
and some analytical results are given also in Section \ref{sec4}.

We first solve the equation:
\begin{equation}
\frac{\partial R^{2}(\tau, \sigma)}{\partial\sigma} = 0
\label{maxim}
\end{equation}
for fixed time $\tau$. This leads to a quartic equation in $\cos (2\sigma)$:
\begin{equation}
\cos^4(2\sigma) + a\cos^3(2\sigma) + b\cos^2(2\sigma) + c\cos(2\sigma) + d = 0
\label{quartic}
\end{equation}
where the coefficients $(a, b, c, d)$ depend on time $\tau$ as well as on
the parameters $(\alpha ,
\beta)$. The explicit expressions are given in the Appendix. The solutions
to equation (\ref{quartic}) are
explicitly known, leading to
$\sigma =
\sigma(\tau)$ for given values of $(\alpha ,
\beta)$. By insertion of these solutions
$\sigma = \sigma(\tau)$ into $R^2(\tau,\,\sigma)$, it is then
straightforward to
obtain the maximal distance in the square bracket of Eq. (\ref{defr}).
This is now
a function of $\tau$, which finally has to be minimized over
$\tau\in\left[0; \pi\right]$.
For more details, see the Appendix and Section \ref{sec4}.

The result is shown in Fig. 1, that is,
the minimal string size $R^2$ as
a function of ($\alpha ,\beta$). Not surprisingly, the minimal string size is
generally of the order 1 (in units of $A$; see Eq. (\ref{turokstring})).
The exception is in
the vicinity of $(\alpha , \beta ) = (0, 0)$, where the minimal string size
is close to
zero. This was also to be expected since the vicinity of $(0, 0)$ describes
the near-spherical strings, and those are the only ones expected to have a
chance to collapse, due to their relatively small angular momentum. These
strings will be considered in detail in Section \ref{sec4}.

However, the computation of the minimal string size $R^2$ gives some more
information about the Turok-strings, namely the time(s) $\tau$ at which the
strings
are minimal. This gives rise to a precise classification of the Turok-strings,
and a subsequent subdivision into 3 different families (see Fig. 2):\\

\noindent
{\bf I}. These strings have their minimal size at $\tau =\pi/2$. That is,
starting
from their original size at $\tau$ = 0, they generally contract to their
minimal size at
$\tau =\pi/2$, and then generally expand back to their original size at
$\tau = \pi$.\\

\noindent
{\bf II}. These strings start from their minimal size at $\tau$ = 0. Then they
generally expand towards their maximal size and then recontract towards their
minimal size at $\tau = \pi$.\\

\noindent
{\bf III}. These strings have their minimal size at two values of $\tau$
symmetrically around
$\pi/2$. That is, they first generally contract and reach the minimal size
at some
$\tau_0\in\left[0;
\pi/2\right]$. Then they expand for a while, and then recontract and reach
the minimal size
again at $\tau =\pi - \tau_0$. Then they expand again towards the original
size
at $\tau = \pi$. In this family of strings, the value of $\tau_0$ depends on
$(\alpha , \beta)$.\\
\\
It must be stressed that the strings in most cases do not expand or
contract {\it isotropically}.
They typically expand in some directions while contracting in other directions.
This is why we use the expressions ``generally expand" and ``generally
contract", which refer to the
minimal string size as a function of time (the radius of the minimal
3-sphere enclosing the
string, as a function of time).

Notice that besides the three above-mentioned families of strings, there
are a number of
degenerate cases at the different boundaries. For instance, at the boundary
$\alpha = 1$, the
strings have their minimal size at $\tau = 0$, $\tau = \pi/3$, $\tau =
2\pi/3$ and $\tau = \pi$.
On the other hand, at the boundary between regions {\bf I} and {\bf II},
the strings have their
minimal size at $\tau = 0$, $\tau = \pi/2$ and $\tau = \pi$. Notice also
that the
3 points $(\alpha , \beta) = (1,0)$, $(\alpha , \beta) = (1,1)$ and
$(\alpha , \beta) = (0,1)$ correspond to rigidly rotating strings, thus they
have their minimal (and maximal) size at $\underline{all}$ times.\\

Let us close this section with a comparison with the result (\ref{polz1}) of
Ref. \cite{pol}.
We see that the result (\ref{polz1}) is correct in the region
{\bf I} of parameter-space,
but incorrect in regions {\bf II} and {\bf III}.\\

\section{Probability of String Collapse}
\label{sec4}
\setcounter{equation}{0}

In this section we consider the question of collapse of the Turok-strings
(\ref{turokstring}). As already discussed
in the previous section, the only strings with
a chance to collapse are those corresponding to parameters
$(\alpha , \beta)$ located
in the vicinity of $(0 , 0)$. That is, we need only consider strings in the
family
{\bf I} of Fig. 2. Using the results of the Appendix,
it is then straightforward to
show that the minimal string size $R^2$, as defined in Eq. (\ref{defr}),
is given by:
\begin{equation}
{R}^2 = \mbox{Max} \left( R^{2}_1,\,R^{2}_2\right)
\label{collapse}
\end{equation}
where:
\begin{equation}
\frac{R^{2}_1}{A^2} = \frac{4{\alpha}^2}{9}
\label{rone}
\end{equation}
\begin{equation}
\frac{R^{2}_2}{A^2} = \left(\sqrt{\alpha\left(1-\alpha\right)} -
\sqrt{\beta\left(1-\beta\right)}\right)^2 + \left(\frac{\alpha}{3} -
\beta\right)^2
\label{rtwo}
\end{equation}
Notice that Eq. (\ref{rtwo}) is precisely the result (\ref{polz2})
of Polnarev and Zembowicz
\cite{pol}. However, in Ref. \cite{pol}, the other solution (\ref{rone}) was
completely
missed, and this is actually the relevant solution in Eq. (\ref{collapse})
in approximately
half of the parameter-space $\left(\alpha , \beta\right)$.

According to the hoop-conjecture (see for instance \cite{hoop}), the
condition for
collapse to a black hole is then given by
Eq. (\ref{condition}), with $R$ given by Eqs.
(\ref{collapse})-(\ref{rtwo}):
\begin{equation}
\mbox{Max} \left( R^{2}_1,\,R^{2}_2\right) \leq\left(4\pi AG\mu\right)^2
\label{newcond}
\end{equation}
which should be solved for $\left(\alpha , \beta\right)$ as a function of
$G\mu$. This can be easily done analytically. The result is shown in
Fig. 3:
the part of parameter-space fulfilling inequality (\ref{newcond})
is bounded by the
$\alpha$-axis, the $\beta$-axis, the straight line $\alpha = 3R_S/2A$ and
the two curves:
\begin{equation}
\beta_{\pm}\left(\alpha\right) = \frac{-16\alpha^{3} - 12\alpha^{2} +
27\alpha - 9\left(2\alpha - 3\right)R^{2}_S/{A^2} \pm 6\sqrt{D}}
{3\left(-32\alpha^2 + 24\alpha + 9\right)}
\end{equation}
where:
\begin{eqnarray}
{D}\equiv -\alpha\left(1 - \alpha\right)\left[8\alpha^2 - 6\alpha
+ 9\left(\frac{R_S}{A} - 1\right)\frac{R_S}{A}\right]
\left[8\alpha^2 - 6\alpha + 9\left(\frac{R_S}{A} +
1\right)\frac{R_S}{A}\right]\nonumber
\end{eqnarray}
Notice also that,
\begin{equation}
\beta_{+}\left(0\right) = \frac{R^{2}_S}{A^2}
\end{equation}
and
\begin{equation}
\beta_{-}\left(\alpha_{0}\right) = 0 \;\; \quad\mbox{for} \;\;\; \alpha_0 =
\frac{9}{16}
\left(1 - \sqrt{1 - \frac{32}{9}\frac{R^{2}_S}{A^2}}\,\right)
\end{equation}
The probability for collapse into black holes is then given by the fraction
$f$:
\begin{eqnarray}
{f} = \int_{R\leq R_S}^{}d\alpha\,d\beta =\int_{0}^{\alpha _0} d\alpha
\int_{0}^{\beta_{+}\left(\alpha\right)} d\beta +
\int_{\alpha_0}^{\frac{3R_S}{2A}} d\alpha
\int_{\beta_{-}\left(\alpha\right)}^{\beta_{+}\left(\alpha\right)}d\beta
\nonumber\\
= \int_{0}^{\alpha_0} \beta_{+}\left(\alpha\right) d\alpha +
\int_{\alpha_0}^{\frac{3R_S}{2A}} \left[\beta_{+}\left(\alpha\right) -
\beta_{-}\left(\alpha\right)\right] d\alpha
\label{prob}
\end{eqnarray}
This equation represents the exact analytical result for the probability of
collapse of the Turok-strings (\ref{turokstring}),
for a given value of $R_S/A = 4\pi G\mu$.
The integrals in (\ref{prob}) are of hyper-elliptic
type \cite{erdelyi}, and not
very enlightening in the general case. However, using that typically
$G\mu\ll 1$ (see \cite{vilenkin,shellard}),
a simple approximation is obtained by
keeping only the
leading order terms:
\begin{eqnarray}
{f} = \frac{12\sqrt{6}}{5}\left(4\pi G\mu\right)^{\frac{5}{2}}
\int_{0}^{1} \frac{t^{2}dt}{\sqrt{1-
t^4}}\,+\,\mathcal{O}\left(\left(G\mu\right)^{\frac{7}{2}}\right)
\nonumber \\ =
\frac{3^{\frac{3}{2}}\left(4\pi\right)^4}{5\,
\Gamma^{2}\left(\frac{1}{4}\right)}
\left(G\mu\right)^{\frac{5}{2}}\,+\,\mathcal{O}
\left(\left(G\mu\right)^{\frac{7}
{2}}\right)
\label{approxprob}
\end{eqnarray}
The result (\ref{approxprob}) is a very good approximation
for $G\mu < 10^{-2}$, thus
for any ``realistic" cosmic strings we conclude:
\begin{equation}
{f} \approx 2\cdot10^{3}\cdot\left(G\mu\right)^{\frac{5}{2}}
\end{equation}
which is our final result of this section.

It should be stressed that we have been using the simplest and most naive
version of the hoop-conjecture: namely, we did not take into account the
angular
momentum of the strings. However, numerical studies \cite{casper} of other
families
of strings showed that inclusion of the angular momentum only leads to minor
changes in the final result, so we expect the same will happen here. It
should also
be stressed that we have neglected a number of other physical effects that
might
change the probability of collapse. These include the gravitational field
of the
string and gravitational radiation.

Finally, as in all other discussions of the probability of string collapse,
we are
faced with the problem that we do not know the measure of integration in
parameter-space.
Thus using another measure in Eq. (\ref{prob}) would generally
give a different
result (see also Ref. \cite{pol}).\\

\section{Conclusion}
\label{sec5}
\setcounter{equation}{0}

In this paper we examined, using purely analytical methods, the question
of collapse of Turok's
two-parameter family of cosmic strings \cite{turok}. We made a complete
classification
of the strings  according to the
specific time(s) the
minimal string size is reached during one period. This revealed that
the previously obtained
results \cite{pol} were only correct in part of the two-dimensional
parameter-space.

We then
obtained an exact analytical expression for the probability $f$ of collapse
into black holes for
the Turok strings, which partly agrees with that of Ref. \cite{pol} in the
sense that
$f\propto(G\mu)^{5/2}$. However, we showed
that there is a large numerical prefactor in the relation. This factor is
of the order
$2000$, and not, as previously stated \cite{pol}, of the order $1$.

One might say that it is perhaps not so important whether the prefactor is
$1$ or $2000$
since the exponent will more or less kill the probability of collapse
anyway. This may
very well be true for ``small" values of $G\mu$ (say, $G\mu \sim 10^{-6}$),
but for
``large" values of $G\mu$ (say, $G\mu \sim 10^{-2}$) the situation is completely
different. In fact, we shall now argue why it is so important to carefully
compute the prefactor: We find that when using our result, a clear picture
is beginning to emerge. Different computations based on different families
of strings (not surprisingly) produce slightly different exponents, but
they also produce completely different prefactors. Importantly, the two go
in the same direction: An exponent larger by $1$ is followed by a
prefactor larger by a factor $100$ (roughly speaking). For instance, the
one-parameter family of Kibble and Turok \cite{kibble} gives $f\approx
16\pi^2(G\mu)^2$, our computation for the two-parameter family gives
$f\approx 2000(G\mu)^{5/2}$ and the Caldwell-Casper computation
\cite{casper} gives $f\approx 10^5(G\mu)^4$. We find it extremely
interesting that for $G\mu\sim10^{-2}$ (which is the range where the
Caldwell-Casper computation is valid), the three computations basically
agree giving $f\approx 10^{-3}-10^{-2}$. We therefore find that our
careful computation of the prefactor is very important and has helped to
understand the discrepancy between previously obtained results. And in
particular, for
``large" values of $G\mu$, there may not even be a discrepancy since the
different exponents and prefactors of the different computations actually
produce more or less the same number for the probability of collapse. For
``small" values of $G\mu$, the picture is unfortunately not so clear, and
more detailed
investigations seem necessary.

It is actually possible to give a physical argument showing that the results of
different computations (if they are done correctly) should merge for
``large" values of
$G\mu$. Consider the rigidly rotating straight string, corresponding to
$\alpha=0$ and $\beta=1$. It is easy to show that it is precisely inside its
Schwarzschild radius for
$G\mu=(4\pi)^{-1}$. Now, it is well-known that the rigidly rotating string
has the
maximal angular momentum per energy (it is exactly on the leading Regge
trajectory), and
therefore is expected to be the most difficult string to collapse into a
black hole.
Therefore, for
$G\mu\approx (4\pi)^{-1}$ all other strings have already collapsed, and we
should expect
$f\approx 1$. This is indeed the case for the three above mentioned
computations, while
it does not hold for the computation of Ref.\cite{pol}. More generally, the
condition
$f((4\pi)^{-1})\approx 1$ can be considered as a physical boundary
condition, and
therefore can be used to immediately rule out some of the previously obtained
results for $f$; for instance those of \cite{pol} and \cite{honma}.

As a possible continuation of our work, it would be very interesting to
consider more
general multi-parameter families of strings, to see how general our
result for $f$ actually is.
Such families of strings have been constructed and considered for instance in
\cite{engle,chen}, and more general ones can be obtained along the lines of
\cite{brown}.

Unfortunately,
there are also still some open questions, as we discussed at the end of Section
\ref{sec4}. The main problem seems to be that we still do not know exactly
what is the
measure of integration in parameter-space
\cite{pol}.
\newpage
\ \vskip 0.3cm
\noindent
{\bf \large Acknowledgements} \\

\noindent
One of us (A.L.L.) would like to thank M.P. Dabrowski for discussions on
this topic, at an early
stage of the work. \\

\appendix
\section*{Appendix: The Minimal String Size}
\setcounter{section}{1}
\setcounter{equation}{0}

In this appendix, we give some details of the results presented in the
sections \ref{sec3} and \ref{sec4}. \\

The distance from origo to the string, as a function of $\tau$ and $\sigma$,
is conveniently written as a polynomial in $\sin\left(2\sigma\right)$ and
$\cos\left(2\sigma\right)$. From (\ref{defdist}) and (\ref{turokstring}):
\begin{eqnarray}
R^2\left(\tau, \sigma \right)& = &
\frac{A^2}{4}\Big\{C_0+\frac{4\alpha\beta}{3}
\left[C_1\sin\left(2\sigma\right)\,
+ \,C_2\cos\left(2\sigma\right)\right]
\nonumber \\[2mm]
& &\hspace*{-6mm}-\frac{4\alpha\beta}{3}
\left[\cos\left(2\tau\right)\cos\left(2\sigma\right) +
\sin\left(2\tau\right)\sin\left(2\sigma\right)\right]\cos\left(2\sigma \right)
\Big\}
\label{polyn}
\end{eqnarray}
where:
\begin{eqnarray}
C_0 & = & 2 - \frac{8\alpha^2}{9} + 2\Big[2\sqrt{\alpha\left(1 - \alpha\right)}
\sqrt{\beta\left(1 - \beta\right)}\nonumber \\
& &\hspace*{25mm}+ \left(1 - \alpha\right)
\left(1 - \beta\right) +
\frac{\alpha\beta}{3}\Big]\cos\left(2\tau\right)\nonumber \\
C_1 & = & \frac{3}{4\alpha\beta}\left[
\frac{2\alpha}{3}\left(1 - \beta\right)\sin\left(4\tau\right) +
\frac{8\alpha}{3}\left(1 -
\alpha\right)\sin\left(2\tau\right)\right] \label{ugly} \\
C_2 & = & \frac{3}{4\alpha\beta}\Bigg[
\frac{2\alpha}{3}\left(1 - \beta\right)\cos\left(4\tau\right) +
\frac{8\alpha}{3}\left(1 - \alpha\right)\cos\left(2\tau\right) \nonumber \\
& &\hspace*{6mm}+ 2\left(2\sqrt{\alpha\left(1 - \alpha\right)}
\sqrt{\beta\left(1 - \beta\right)}-\beta\left(1-\alpha\right)\right)\Bigg]
\nonumber
\end{eqnarray}
Then it is straightforward to show that the condition (\ref{maxim}) leads to
\begin{equation}
x^4 + ax^3 + bx^2 + cx + d = 0
\label{quarticagain}
\end{equation}
where $x \equiv \cos\left(2\sigma\right)$ and:
\begin{eqnarray}
a & = & -C_1\sin\left(2\tau\right) - C_2\cos\left(2\tau\right)\nonumber \\
b & = & \frac{C^{2}_1 + C^{2}_2}{4} -1 \\
c & = & \frac{C_1}{2}\sin\left(2\tau\right) +
C_2\cos\left(2\tau\right)\nonumber \\
d & = & \frac{\sin^2\left(2\tau\right) - C^{2}_2}{4}\nonumber
\end{eqnarray}
The solutions to Eq. (\ref{quarticagain}) can be written down in closed form.
Define
\begin{eqnarray}
{\mathcal X} & \equiv & b^2 - 3ac +12d\nonumber \\
{\mathcal Y} & \equiv & 2b^3 - 9abc + 27c^2 + 27a^2d - 72bd\\
{\mathcal Z} & \equiv & \left[{\mathcal Y} + \sqrt{-4 {\mathcal X}^3 +
{\mathcal Y}^2}\right]^\frac{1}{3}\nonumber \\
{\mathcal W} & \equiv & \frac{2^{\frac{1}{3}}\mathcal{X}}{3\mathcal{Z}} +
\frac{\mathcal{Z}}{3 \: 2^{\frac{1}{3}}}\nonumber
\end{eqnarray}
Then the 4 solutions are:
\begin{eqnarray}
x_{1,\,2} = -\frac{a}{4} - \frac{1}{2}\sqrt{\frac{a^2}{4} - \frac{2b}{3}
+{\mathcal W}} \mp
\frac{1}{2}\sqrt{\frac{a^2}{2} - \frac{4b}{3} - {\mathcal W} -
\frac{\left(-a^3+4ab-8c\right)}{4\sqrt{\frac{a^2}{4} - \frac{2b}{3}
+{\mathcal W}}}}\nonumber \\
x_{3,\,4} = -\frac{a}{4} + \frac{1}{2}\sqrt{\frac{a^2}{4} - \frac{2b}{3}
+{\mathcal W}} \mp
\frac{1}{2}\sqrt{\frac{a^2}{2} - \frac{4b}{3} - {\mathcal W} +
\frac{\left(-a^3+4ab-8c\right)}{4\sqrt{\frac{a^2}{4} - \frac{2b}{3}
+{\mathcal W}}}}\nonumber \\
\end{eqnarray}
which give $\sigma = \sigma\left(\tau\right)$ for given values of
$\left(\alpha , \beta\right)$. These solutions are inserted
into (\ref{polyn}) and then
the result of the square bracket in Eq. (\ref{defr}) is determined. Finally
this
function
of $\tau$ must be minimized.\\

As an example, consider strings in the region {\bf I} of parameter-space;
see Fig. \ref{fig2}.
Using the above formulas, one finds:
\begin{eqnarray}
x_{1,\,2} & = & -\frac{C_2}{2}\nonumber \\
x_{3,\,4} & = & \mp\,1
\end{eqnarray}
with
\begin{equation}
C_2 = \frac{3}{2\alpha\beta}\left[\left(\frac{\alpha}{3}+\beta\right)
\left(\alpha -\beta\right)- \left(\sqrt{\alpha\left(1 - \alpha\right)} -
\sqrt{\beta\left(1 - \beta\right)}\,\right)^2\right]
\end{equation}
Insertion into Eq. (\ref{polyn}) then leads directly
to the result for the minimal
string size in the region {\bf I},
\begin{equation}
{R}^2 =\mbox{Max}\left(R^{2}_{1},\,R^{2}_{2}\right)
\end{equation}
where
\begin{eqnarray}
R^{2}_{1} & = & \frac{4A^{2}\alpha^2}{9}\nonumber \\
R^{2}_{2} & = & A^{2}\left[\left(\sqrt{\alpha\left(1 - \alpha\right)} -
\sqrt{\beta\left(1 - \beta\right)}\,\right)^{2} +
\left(\frac{\alpha}{3}-\beta\right)^{2}\right]
\end{eqnarray}
c.f. Eqs. (\ref{rone})-(\ref{rtwo}).\\

\newpage
\centerline{\Large CAPTIONS FOR FIGURES} \vskip 2cm

\noindent Figure 1: The radius of the minimal 3-sphere completely enclosing a 
string with
parameters $(\alpha, \beta)$ plotted for all parameter-space. 
Notice that $R^2$
is close to zero only near $(\alpha, \beta) = (0,0)$, 
that is for the near-spherical string configurations. \vskip 1cm

\noindent Figure 2: The considered strings fall into three families. The ones
that reach their minimal size $R$ at $\tau = \pi /2$ ({\bf I}), 
at $\tau = 0,\, \pi$ ({\bf II}) and 
at $\tau = \tau _0 ,\, \pi - \tau _0$ 
for $\tau _0 \in ]0,\pi /2[$ ({\bf III}). \vskip 1cm

\noindent Figure 3: The region of parameter-space which contains the 
strings falling inside
their own Schwarzschild radius is bounded by $\alpha = 0$, $\beta = 0$, $\alpha
=3R_S/2A$ and the two curves $\beta _\pm$. Here $G \mu = 10^{-2}$ is chosen in 
order to illustrate the general form of the region. As $G \mu$ decreases, the
region is relatively stretched out and becomes quite narrow. \vskip 1cm

\newpage

\begin{figure}[htb]
\vspace{-6cm}
\centerline{\psfig{file=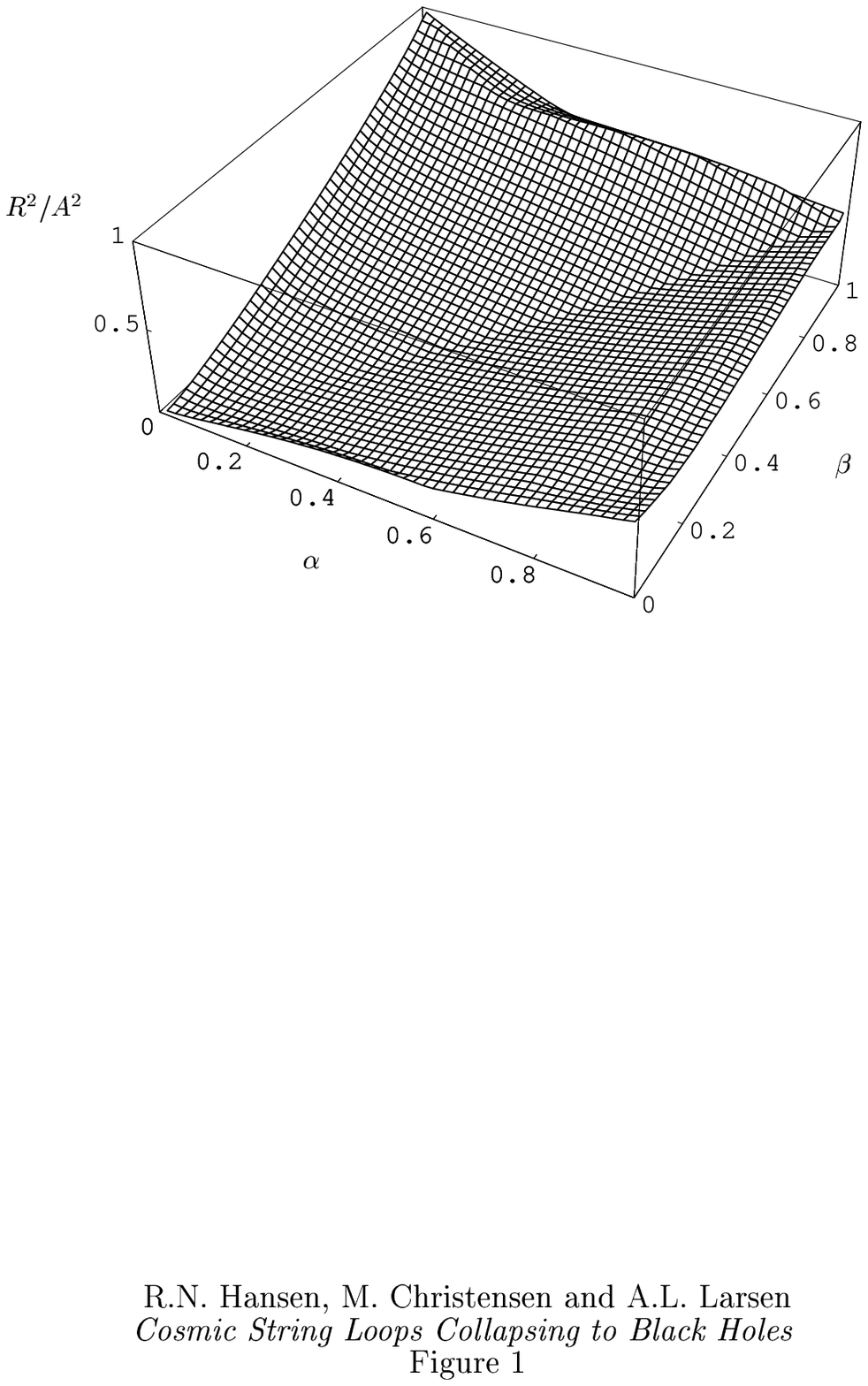,height=32cm}}
\label{fig1}
\end{figure}

\begin{figure}[htb]
\vspace{-6cm}
\centerline{\psfig{file=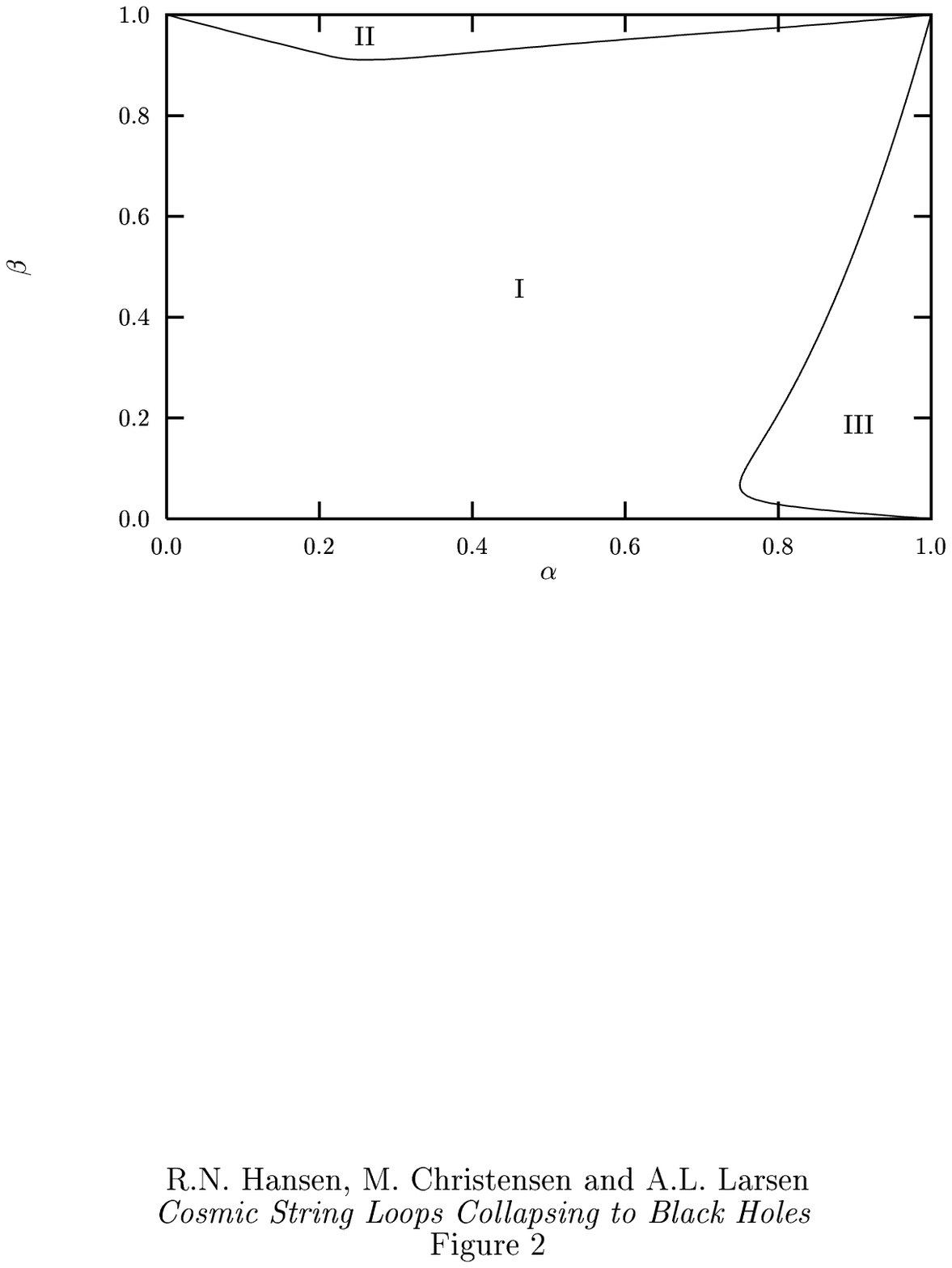,height=32cm}}
\label{fig2}
\end{figure}

\begin{figure}[htb]
\vspace{-6cm}
\centerline{\psfig{file=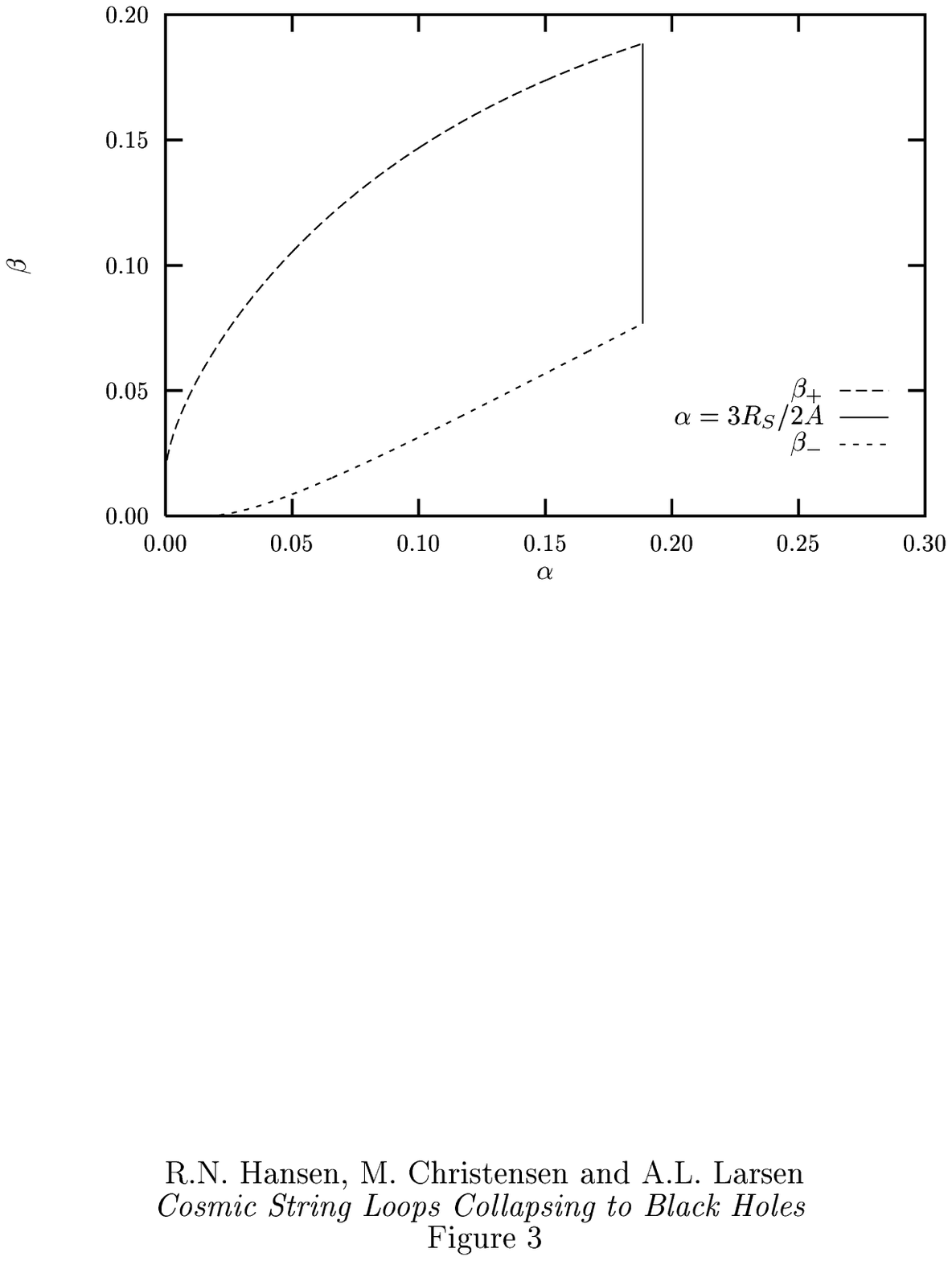,height=32cm}}
\label{fig3}
\end{figure}

\end{document}